\documentclass[aps,pre,preprint,superscriptaddress]{revtex4-1}

\usepackage{graphics}
\usepackage{color}
\usepackage{amsmath, amssymb}
\usepackage{type1cm}
\usepackage{mathrsfs}
\usepackage{url}

\begin{document}


\title{Smoothing effect for spatially distributed renewable resources and its impact on power grid robustness}


\author{Motoki Nagata}
\affiliation{Graduate School of Information Science and Technology, The University of Tokyo, Tokyo, Japan}
\author{Yoshito Hirata}
\affiliation{Graduate School of Information Science and Technology, The University of Tokyo, Tokyo, Japan}
\affiliation{Institute of Industrial Science, The University of Tokyo, Tokyo, Japan}
\affiliation{CREST, JST, Saitama, Japan}
\author{Naoya Fujiwara}
\affiliation{Center for Spatial Information Science, The University of Tokyo, Chiba, Japan}
\author{Gouhei Tanaka}
\affiliation{Graduate School of Information Science and Technology, The University of Tokyo, Tokyo, Japan}
\affiliation{Institute of Industrial Science, The University of Tokyo, Tokyo, Japan}
\affiliation{Graduate School of Engineering, The University of Tokyo, Tokyo, Japan}
\author{Hideyuki Suzuki}
\affiliation{Graduate School of Information Science and Technology, The University of Tokyo, Tokyo, Japan}
\affiliation{CREST, JST, Saitama, Japan}
\author{Kazuyuki Aihara}
\affiliation{Graduate School of Information Science and Technology, The University of Tokyo, Tokyo, Japan}
\affiliation{Institute of Industrial Science, The University of Tokyo, Tokyo, Japan}
\affiliation{CREST, JST, Saitama, Japan}
\affiliation{Graduate School of Engineering, The University of Tokyo, Tokyo, Japan}



\date{\today}

\begin{abstract}
In this paper, we show that spatial correlation of renewable energy outputs greatly influences the robustness of power grids.
First, we propose a new index for the spatial correlation among renewable energy outputs.
We find that the spatial correlation of renewable energy outputs in a short time-scale is as weak as that caused by independent random variables and that in a long time-scale is 
as strong as that under perfect synchronization.
Then, by employing the topology of the power grid in eastern Japan,
we analyze the robustness of the power grid with spatial correlation of renewable energy outputs. The analysis is performed by using a realistic differential-algebraic equations model and the result shows that the spatial correlation of the energy resources strongly degrades 
the robustness of the power grid.
Our result suggests that the spatial correlation of the renewable energy outputs should be taken into account when estimating the stability of power grids.
\end{abstract}


\maketitle

\section{INTRODUCTION}
A lot of renewable energy resources are being incorporated into power networks for realizing low carbon societies.
More and more renewable energy resources will be introduced in all over the world.
For example, in Japan, 
the amount of PV introduction in 2030 is expected to be about ten times larger than that in 2012
\cite{Renewable}.
The large amount of renewable energy resources can cause large power fluctuations that may result in blackouts.
When analyzing the robustness of power grids based on mathematical models, linearization schemes are not available in the presence of large power fluctuations, because the model behavior is far from a steady state.
A method that enables to quantify the robustness of stable states against large fluctuations in differential equations was proposed in Ref.\ \cite{Menck2013}, where the robustness is measured by the volume of the basins of attraction of the stable states. This method was used to investigate the relation between the network topology and the robustness of power grids against large fluctuations \cite{Menck2014}.

In the studies of power grids in the physics community, 
the Kuramoto-like phase oscillator model, which qualitatively corresponds to the swing equation in the electrical engineering community, has often been used in order to clarify various 
properties of power grids \cite{Menck2014,Filatrella2007,Buzna2009,Lozano2012,Rohden2012,Witthaut2012,Dorfler2012,Nagata2013}.
Although the model is simple and mathematically tractable, it
has drawbacks 
that it does not take into account  some factors which play important roles in the stability of power grids, particularly fluctuations in the voltage amplitude and the adjustment of reactive power in loads \cite{Nagata2014}. In order to study the effects of these factors on the stability of the power grids with relatively simple models, a new mathematical model governed by a differential-algebraic equations was proposed \cite{Sakaguchi2012}.
In this model,
the properties of generators are described using the swing equations
\cite{Filatrella2007,Buzna2009,Lozano2012,Rohden2012,Witthaut2012,Dorfler2012,Nagata2013},
while the properties of loads (or substations) are determined with
the power flow calculation.
We proposed a new method to analyze the robustness against large fluctuations 
in the power grid system governed by
this differential-algebraic equations model
in Ref.\ \cite{Nagata2014}.

Among various aspects of renewable energy outputs intensively studied so far \cite{Lei2009},
the smoothing effect
is notable \cite{Nagoya2011,Nagoya2013}.
The smoothing effect indicates that the fluctuations in the total amount of renewable energy outputs
are smaller than the sum of fluctuations in the individual renewable energy outputs  
because the individual fluctuations are balanced out.
Although the smoothing effect seems to enhance the robustness of the power grid, we need to take into account the spatial correlation of the weather conditions \cite{Milan2013,Lind2014} which may destabilize the system locally. Therefore, it is significant to understand the relation between the spatial correlation and the robustness of power grids.

In this study, we propose a new index that quantifies the spatial correlation of renewable energy resources, for characterizing the smoothing effect.
In addition, we analyze the relation between the smoothing effect and 
the robustness of power grids by extending the method proposed in Ref.\ \cite{Nagata2014}. 
We show that the spatial correlation, and thus the smoothing effect makes power grids more fragile.

This paper is organized as follows. First, in Sec.\ II,
we give a brief overview of the model of the power grids.
Then, in Sec.\ III, we propose a new index for estimating the spatial correlation
of renewable energy outputs and calculate the index value from real data.
Section IV is devoted to clarifying the relation
between the smoothing effect and the robustness of power grids
using the real power grid topology in eastern Japan.
Finally, in Sec.\ V, we conclude this paper. 
\section{A DYNAMICAL MODEL OF POWER GRIDS AND A ROBUSTNESS MEASURE}
\subsection{Differential-algebraic equations model}
We introduce the differential-algebraic equations model for power grids proposed by Sakaguchi and
Matsuo \cite{Sakaguchi2012}. Detailed analysis was given in Ref.\ \cite{Nagata2014}.

Power grids consist of generators, loads, and end-users including buildings, factories, and 
houses.
Recently, an increasing number of solar panels are connected to the end-users.
The transmitted energy from generators to loads is equal to 
the difference between the energy consumed by end-users and 
the output energy of the solar panels.  
Therefore, large fluctuations in powers generated by the solar panels can cause 
large fluctuations of the transmitted energy from generators to loads.
The voltages for the transmission lines between generators and loads are very high
(500 kV or 275 kV in Japan).
On the other hand, the voltages for the transmission lines between loads and end-users
are relatively low (66 kV or 6.6 kV in Japan).
Since the voltages are converted from high values to low values at substations,
substations can be seen as loads from the high-voltage power grid.
In this study, we focus on the parts with high-voltage transmission lines, i.e., 
generators and loads.
The numbers of generators, loads, and branching points are
denoted by $N_g$, $N_l$, and $N_b$, respectively, and 
the total number of generators and loads by $N_r \equiv N_g + N_l$. 

Power grids are regarded as complex networks consisting of nodes which correspond to 
generators, loads, and branching points.
We denote the voltage of node $j$ by $V_j=|V_j| \exp(\mathrm{i}\theta_j)$ for $j=1, \ldots, N_r$, where
$|V_j|$ and $\theta_j$ are the amplitude and the phase of the voltage, respectively.
Complex numbers are useful to represent the behavior of alternate-current voltages.
The resistance is much smaller than the reactance in the transmission lines, and therefore, we can consider a power network where the resistance is neglected, i.e., a lossless network.
We use the Kron's reduction in the same way as Refs. \cite{Kundur1994,Machowski2008,Nagata2014} and denote the reduced susceptance between node $j$ and node $k$ by $B_{jk}$.

We regard the generators as phase oscillators.
We denote the difference between the phase $\theta_j$ and the phase advance by the standard frequency $\Omega$ by $\phi_j \equiv \theta_j-\Omega t$.
This standard frequency $\Omega$ is $2\pi\times 60$ s$^{-1}$ in the United States and western Japan, and
$2\pi\times 50$ s$^{-1}$ in Europe and eastern Japan. 
The behavior of each generator is described as follows \cite{Filatrella2007}:
\begin{equation}
M_j\ddot{\phi}_j+D_j\dot{\phi}_j=P_{j}^m-\sum_{k=1}^{N_r} |V_j||V_k|B_{jk}\sin(\phi_j-\phi_k),
\label{eq:swingequation}
\end{equation}
where $M_j$ is the inertial coefficient, $D_j$ is the damping coefficient, and
$P_j^m$ is the mechanical input energy.
Equation (\ref{eq:swingequation}) is called  the `swing equation' and derived from 
the energy conservation law.
We adopt the local feedback control as follows \cite{Sakaguchi2012}:
\begin{equation}
\dot{P}_{j}^m=-K_j\dot{\phi}_j,\label{eq:fb}
\end{equation}
where $K_j$ corresponds to the inverse of the time constant.

The behavior of each load
is written as follows:
\begin{eqnarray}
P_j=\sum_{k=1}^{N_r} |V_j||V_k|B_{jk}\sin(\phi_k-\phi_j), \label{eq:Piload}\\
Q_j=\sum_{k=1}^{N_r} |V_j||V_k|B_{jk}\cos(\phi_k-\phi_j), \label{eq:Qiload}
\end{eqnarray}
where $P_j$ and $Q_j$ are the consumed effective energy and the consumed reactive energy in loads, respectively.
If the reactive energy is supplied in loads, $Q_j$ is negative.
Equations (\ref{eq:Piload}) and (\ref{eq:Qiload}) are derived from Kirchhoff's and Ohm's laws.
Each iteration of our numerical simulation consists of two steps:
First, we used
the fourth-order Runge--Kutta method to simulate the generators'
Eq. (\ref{eq:swingequation}). We set the time step to $0.005$.
Second, we used Newton's method for the loads' Eqs.
(\ref{eq:Piload}) and
(\ref{eq:Qiload}).
We repeated this iteration for sufficiently many times until a convergence is reached.
We use the values of $|V_j|$ and $\phi_j$ at load nodes in the previous step
to obtain $\phi_j$ of generator node $j$ when simulating the generators'
Eq. (\ref{eq:swingequation}).
On the other hand,
we use the value of $\phi_j$ at generator $j$ in the current step 
to obtain $P_j$ and $Q_j$ of load node $j$ when solving the loads'
Eqs. (\ref{eq:Piload}) and
(\ref{eq:Qiload}).
\subsection{Robustness evaluation}
We set the effective power and reactive power in loads stochastically.
By simulating the generators' Eq. (\ref{eq:swingequation})
and the loads' Eqs.
(\ref{eq:Piload}) and
(\ref{eq:Qiload}), we judge whether the solution for 
Eqs.
(\ref{eq:Piload}) and
(\ref{eq:Qiload})
exists.
The case where the solution does not exist corresponds to the power shortage,
because this absence means that the power grid cannot realize such a power flow.
In this case, the voltage collapse occurs.
We calculate the probability that the solution exists and
call this probability the stability rate. We evaluate the robustness based on the stability rate.
\section{SPATIAL CORRELATION OF RENEWABLE ENERGY OUTPUTS}
\subsection{Smoothing effect}
The smoothing effect
 \cite{Nagoya2011,Nagoya2013}, that
the fluctuations in the total of renewable energy outputs
are smaller than the sum of the fluctuations in the individual outputs,
 is a remarkable aspect of renewable energy outputs.
We formulate the smoothing index in the next subsection.
\subsection{Smoothing index}
Let $r_j(t)$ and $N$  be the power output in node $j$ at time $t$ and the number of nodes, respectively.
In addition, we denote 
the sample standard deviation of the component of a row vector
$\bold{X}$ by $SD(\bold{X})$. 
We quantify the strength of the smoothing effect for time window size $W$ using the following smoothing index:
\begin{eqnarray}
SI(W)=\frac
{\displaystyle\sum_{j=1}^N SD\biggl(\Bigl(r_j(W+1), r_j(W+2), \ldots, r_j(T)\Bigr) -
\Bigl(r_j(1), r_j(2), \ldots, r_j(T-W)\Bigr)\biggr)}
{SD\biggl(\displaystyle\sum_{j=1}^N \Bigl(r_j(W+1), r_j(W+2), \ldots, r_j(T)\Bigr) -
\displaystyle\sum_{j=1}^N \Bigl(r_j(1), r_j(2), \ldots, r_j(T-W)\Bigr)\biggr)}
,\nonumber
\\
\label{eq:si}\end{eqnarray}  
where the total number of time steps is expressed as $T (>W)$.
The smoothing index defined by Eq. (\ref{eq:si}) represents the correlation among nodes.
If the correlation is zero, i.e., the fluctuations of noise balance out, 
and if probability distribution of all nodes is the same,
the smoothing index becomes $\sqrt{N}$.
On the other hand, if $r_j(t)$ is positively correlated with each other,  
the smoothing index becomes lower than $\sqrt{N}$.
If the Pearson correlation coefficient (PCC) between each two nodes is 1,
then the smoothing index is 1.
The more the smoothing index is, the stronger the smoothing effect is.

\subsection{Smoothing index of renewable energy output data}
\begin{figure}[h]
\vspace{0cm}
\begin{center}
\scalebox{0.4}{
\includegraphics{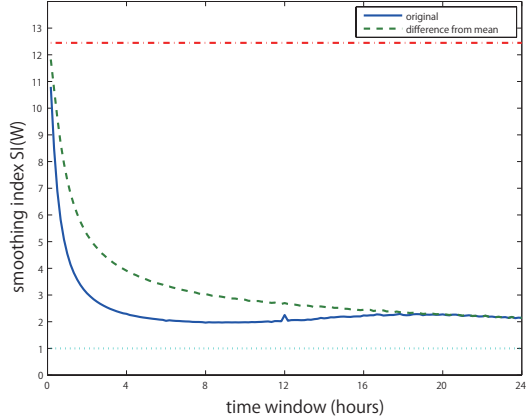}
}
\caption{The smoothing index of wind speed data at 155 measurement points in Japan for different size of time window.
The horizontal axis represents the size of the time window. The step size is set at 10 minutes.
The vertical axis represents the smoothing index.
The dash-dotted line, indicating $SI=\sqrt{155} \approx 12.4$, represents the case where 
the PCC of each two nodes is zero, i.e., where the fluctuations of noise are balanced out.
The dotted line, indicating $SI =1$, represents the case where the PCC of each two nodes is 1.
The solid line represents the smoothing index for the real data (see the main text for details). 
The dashed line represents the smoothing index for the deviation of the data from its mean.
\label{fig:si}}
\end{center}
\end{figure}
\begin{figure}[h]
\vspace{0cm}
\begin{center}
\scalebox{0.5}{
\includegraphics{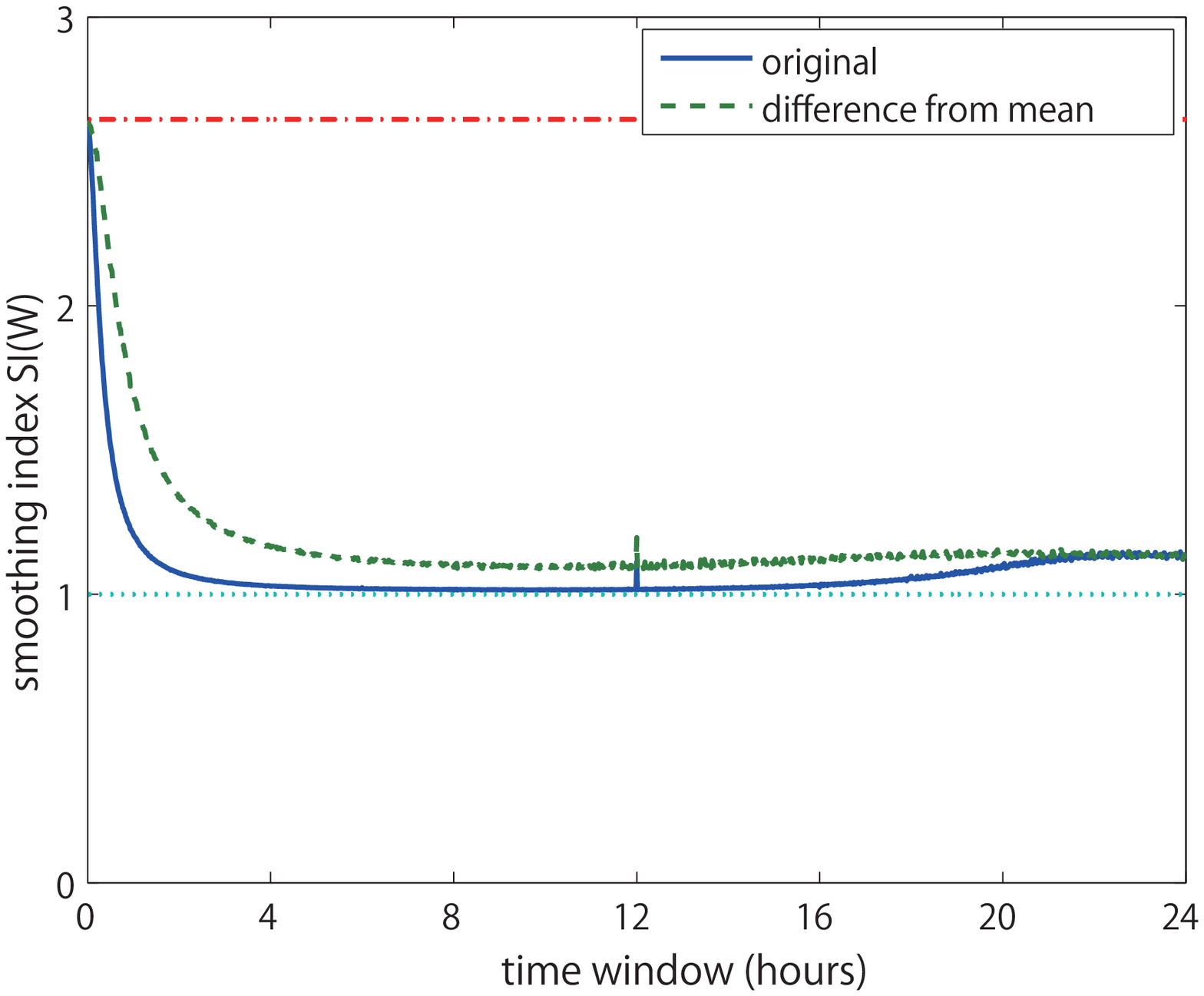}
}
\caption{The smoothing index for solar radiation data at 7 measurement points in Japan for different size of the time window.
The horizontal axis represents the size of the time window.
The step size is 1 minute.
The vertical axis represents the smoothing index.
The dash-dotted line, indicating $SI=\sqrt{7}$, represents the case where 
the PCC of each two nodes is zero, i.e., where the fluctuations of noise are balanced out.
The dotted line, indicating $SI=1$, represents the case where the PCC of each two nodes is 1.
The solid line represents the smoothing index for the real data.
The dashed line represents the smoothing index for the deviation of the data from its mean.
In the solar radiation data, there exist corrupted data due to the breakdown or checkback of
measurement equipments.
We excluded such data when calculating the mean and the smoothing index.
\label{fig:sisolar}}
\end{center}
\end{figure}
We calculated the smoothing index for wind speed data at 155 measurement points in Japan (Fig. \ref{fig:si}) and
that for solar radiation data at 7 measurement points in the Kanto area, Japan (Fig. \ref{fig:sisolar}) for different size of the time window.
The time period for wind speed data and that for solar radiation data
is between January 1 2010 and December 31 2012 and
between January 1 2011 and December 31 2011, respectively.
As shown in these figures, the smoothing index of these data takes the value between 1 and $\sqrt{\mathrm{\# points}}$.
The dash-dotted line represents the case where 
the PCC of each two nodes is zero, i.e., where the fluctuations of noise are cancelled out.
If we assume the independence of the fluctuations at two points, we obtain
 $SI=\sqrt{\mathrm{\#points}}$.
The dotted line corresponding to $SI=1$ represents the case where the PCC of each two nodes is 1.
The solid line represents the smoothing index for the real data.
In the region of $W\le 12$, the smoothing index decreases monotonically.
This is because the larger the time window size is,
the more similar the temporal changes at the individual points are, independently of their locations  within Japan.
This spatially independent trend is attributed to the nationwide weather.
On the other hand, in the region of $W\ge 12$, the smoothing index increases monotonically. 
This is because the effect of the daily periodicity is highest when $W=12$ and
monotonically decreases when $W\ge 12$. 
The output data obtained at the same time each day varies from day to day.
In order to consider this day-by-day variability, we obtained the 24-hour data by calculating the mean of the output data recorded at the same time each day over the days of observation.
The deviation of the output data from its mean is represented by the difference between the output data of each day and the 24-hour data.
The dashed line represents the smoothing index of the deviation of the output data.
The difference between the solid line and the dashed line represents the contribution of the daily periodicity.
The deviation of the output data from its mean is more representative of the smoothing effect.

In summary, the smoothing index of the renewable energy outputs decreases in time. This implies  that the renewable energy outputs of different points fluctuate  independently in a short time scale, and the dependence is stronger if we consider a longer time window.
This result suggests the necessity to study stability of power grids 
taking into account the spatial correlation of the renewable energy outputs.
In the next section, we introduce a spatial correlation of the effective power of different sites in the model and study dependence of the stability of the power grid on the spatial correlation.

\section{Impact of smoothing effect on robustness of the power grids}
We simulated the differential-algebraic equations model for the power grid in eastern Japan by taking into account the smoothing
effect found in the previous section.
\subsection{Eastern Japan power grid}
\begin{figure}[h]
\vspace{-1cm}
\begin{center}
\scalebox{0.5}{
\includegraphics{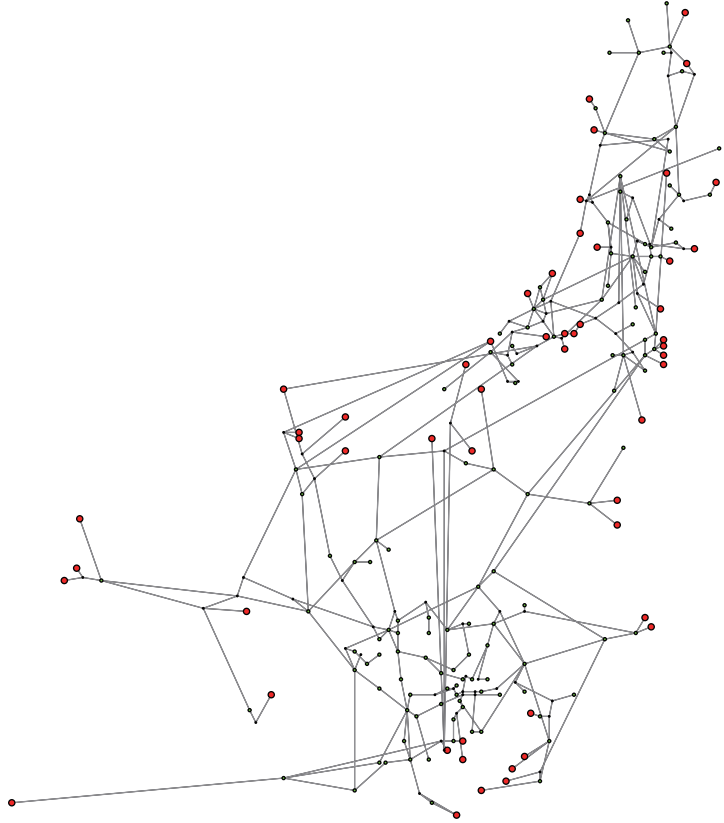}
}
\vspace{0cm}
\caption{(Color online) The topology of the high voltage power grid in eastern Japan that we derived. 
The red circles represent generators, 
the green circles represent loads, and
the small gray circles represent branching points.
\label{fig:eastjapanwithoutlabelchange}}
\end{center}
\end{figure}
Figure \ref{fig:eastjapanwithoutlabelchange} shows the topology of the high voltage power grid in eastern Japan.
It was derived from the reports released by 
Tokyo Electric Power Company Inc. \cite{Tokyo} and
Tohoku Electric Power Company Inc. \cite{Tohoku}.
The figure was produced with the graph visualization software Gephi \cite{Gephi}.
This power grid consists of $N_g=53$, $N_l=126$, and $N_b=69$ nodes.
The number of nodes corresponding to generators and loads is $N_r=N_g+N_l=179$ 
and the total number of nodes in the power grid is $N=248$.
In addition, we set the parameter values to $M_j=1$, $D_j=1$, $K_j=30$, $P_{j}^m(0)=1$, and $Q_j=0$.
The admittance matrix for the links was set as follows \cite{Nagata2014}:
\begin{equation}
B_{jk}=
\left\{
\begin{array}{ll}
10 , &\quad \mbox{if node $j$ and node $k$ are connected},  \\
0 , &\quad \mbox{otherwise}.
\end{array}
\right.
\end{equation}
\subsection{Results}
\begin{figure}[h]
\vspace{0cm}
\begin{center}
\scalebox{0.45}{
\includegraphics{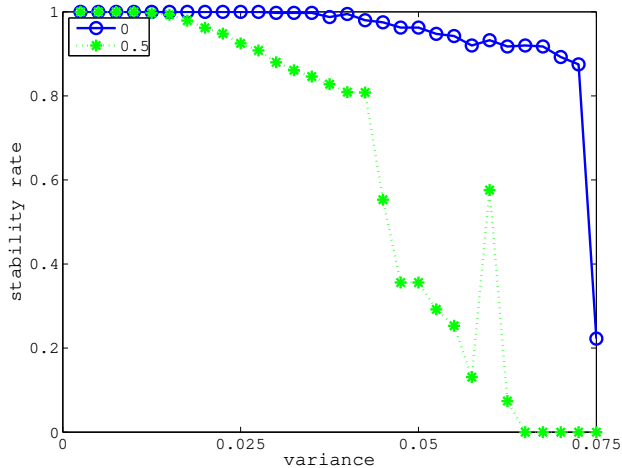}
}
\caption{The relation between the variance of fluctuations and 
the stability rate.
The horizontal axis represents the variance of the fluctuations in each node.
The vertical axis represents the stability rate.
The open circles with the solid line show the result when the PCC of each two nodes
is 0.
The asterisks with the dashed line correspond to the result when the PCC of each two nodes is set to 0.5.
\label{fig:starateeachfluc}}
\end{center}
\end{figure}
We give a fluctuation of consumed effective power in each load, i.e., $P_j$. The mean of the effective power $P_j$ is set to 0.45. In addition, we vary the variance of $P_j$ from 0 to 0.15.
For simplicity, we assume that the mean and the variance of the fluctuations are identical for all nodes and the PCC of each two nodes is the same.
Further, we assume that the autocorrelation of the fluctuations in each node is 0.
Figure \ref{fig:starateeachfluc} shows the relation between the variance of fluctuations and 
the stability rate.
The open circles with the solid line indicate the result when the PCC of each two nodes
is 0.
The asterisks with the dashed line indicate the result when the PCCs are set at 0.5.
Clearly, the power grids become more unstable when the fluctuations are more correlated.
\begin{figure}[h]
\vspace{0cm}
\begin{center}
\scalebox{0.45}{
\includegraphics{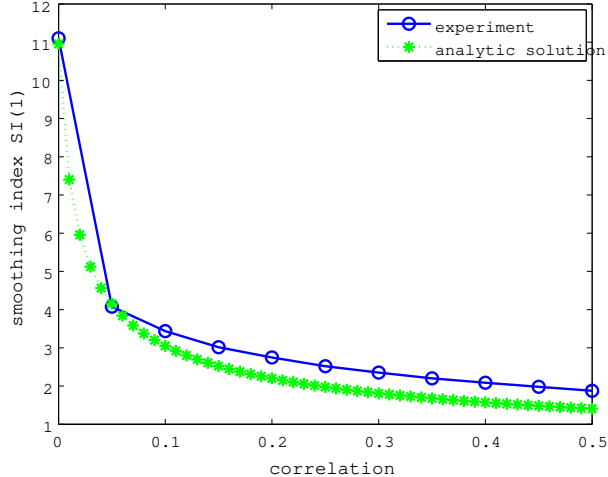}
}
\caption{The relation between the PCC of fluctuations between each two nodes and 
the smoothing index.
The vertical axis represents the smoothing index defined in Eq. (\ref{eq:si}).
The time window size $W$ is equal to one step of the numerical simulation.
The horizontal axis represents the PCC of each two nodes.
The open circles with solid lines show the experimental result.
The asterisks with dashed lines show the analytical solution, which is derived in Appendix.
It is in good agreement with the experiment result.
\label{fig:siandcor}}
\end{center}
\end{figure}

We further fixed the variance of fluctuations in each node and varied the PCC of those between each two nodes.
Figure \ref{fig:siandcor} shows the relation between 
the PCC of each two nodes and 
the smoothing index.
The vertical axis represents the smoothing index defined in Eq. (\ref{eq:si}).
The time window size $W$ is set to be equal to one step of the numerical simulation.
The horizontal axis represents the PCC of fluctuations between each two nodes.
The larger the PCC is, the smaller the smoothing index is.
When the output of each node is independent, the smoothing index is $\sqrt{N_l}=\sqrt{120}$.
In fact, the smoothing index is near $\sqrt{120}$ when the PCC is 0.
\begin{figure}[h]
\vspace{0cm}
\begin{center}
\scalebox{0.45}{
\includegraphics{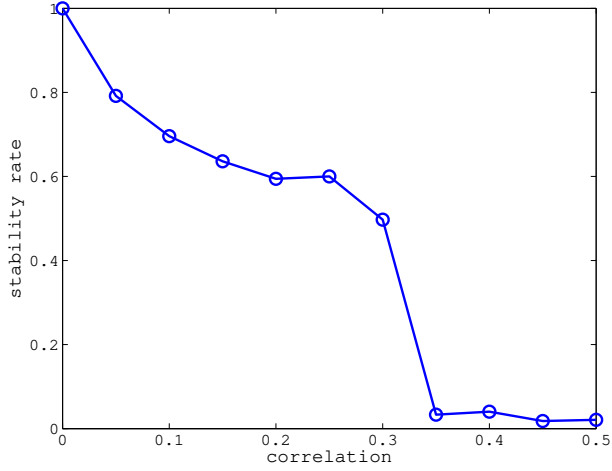}
}
\caption{The relation between the PCC of fluctuations and 
the stability rate.
The horizontal axis represents the PCC of each two nodes.
The vertical axis represents the stability rate.
\label{fig:starate}}
\end{center}
\end{figure}
Figure \ref{fig:starate} shows the relation between the PCCs and the stability rate.
The horizontal axis represents the PCC of each two nodes.
The vertical axis represents the stability rate.
The larger the PCC is, the less stable the power grid is.
\begin{figure}[h]
\vspace{0cm}
\begin{center}
\scalebox{0.45}{
\includegraphics{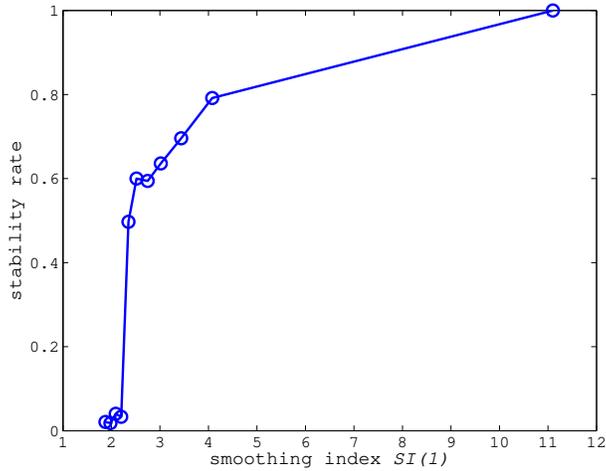}
}
\caption{The relation between the smoothing index and 
the stability rate.
The horizontal axis represents the smoothing index defined in Eq. (\ref{eq:si}).
The time window size $W$ was set to be equal to one step of the numerical simulation.
The vertical axis represents the stability rate.
\label{fig:siandstarate}}
\end{center}
\end{figure}
Figure \ref{fig:siandstarate} shows the relation between the smoothing index and the stability rate.
The horizontal axis represents the smoothing index defined in Eq. (\ref{eq:si}).
The time window size $W$ was set to be one step of the numerical simulation.
The vertical axis represents the stability rate.
The larger the smoothing index is, the more stable the power grid is.
These results mean that the smoothing effect can help stabilize the power grid.
\section{Conclusions}
In this paper, we have proposed a new index, called the smoothing index, 
which quantifies the spatial correlation among renewable energy outputs.
We have analyzed the relation between the smoothing effect and 
the robustness of power grids by using
the mathematical model governed by differential-algebraic equations proposed in Ref.\ \cite{Sakaguchi2012} and the real topology of the power grid in eastern Japan.
We have clarified that the smoothing effect facilitates the stability of the power grid.
In other words, the spatial correlation of the renewable energy outputs degrades the robustness of the power grid, which is an important factor for studying the introduction of the renewable energy.

\begin{acknowledgments}
We thank the Japan Meteorological Agency for providing the wind and solar irradiation datasets used in this study.
This research was supported by Core Research for Evolutional Science and Technology (CREST), Japan Science and Technology Agency (JST).
\end{acknowledgments}

\section*{Appendix: Derivation of the relation between the smoothing index and the PCC}
For simplicity, we assume in the simulation that the mean and the variance of fluctuations are identical for all the nodes,
the PCC between any pair of two nodes is the same, 
and the autocorrelation of each node is 0. 
In the approximate derivation of the relation between the smoothing index and the PCC,
we assume that the sample standard deviation is equal to the standard deviation and the correlation between $r_j(t+1)$ and $r_k(t)$ is 0 for simplicity.
By setting the variance $V(r_j(t))$ to $\sigma^2$ and the correlation of each node to $c$,
the smoothing index is calculated as follows:
\begin{eqnarray}
SI(1)&=&\frac
{\sum_{j=1}^N sd(r_j(t+1)-r_j(t))}
{sd(\sum_{j=1}^N (r_j(t+1)-r_j(t)))} \nonumber \\
&=&\frac
{\sum_{j=1}^N \sqrt{V(r_j(t+1)-r_j(t))}}
{\sqrt{V(\sum_{j=1}^N (r_j(t+1)-r_j(t)))}} \nonumber \\
&=&\frac
{N \sqrt{(V(r_j(t+1))+V(r_j(t)))}}
{\sqrt{V(\sum_{j=1}^N r_j(t+1))+V(\sum_{j=1}^N r_j(t))}} \nonumber \\
&=&\frac
{N\sqrt{2\sigma^2}}
{\sqrt{2N\sigma^2+2N(N-1)c\sigma^2}} \nonumber \\
&=&\frac
{\sqrt{N}}
{\sqrt{1+(N-1)c}}.
\end{eqnarray}
Therefore, $SI(1)=\sqrt{N}$ if $c=0$ and 
$SI(1)=1$ if $c=1$.
The analytic solution of the relation between the smoothing index $SI(1)$ and the correlation is depicted for real data in 
Fig. \ref{fig:siandcor}. It is in good agreement with the experimental result.

\bibliography{book5withnumber.bib}

\end{document}